\documentclass[aps,prl,twocolumn,superscriptaddress]{revtex4}
\begin{document}

\title{IS THERE A DARK MATTER SIGNAL IN THE GALACTIC POSITRON ANNIHILATION RADIATION ?}

\author{R. E. Lingenfelter}
\affiliation{Center for Astrophysics and Space Sciences,
University of California San Diego, La Jolla, CA 92093}
\author{J. C. Higdon}
\affiliation{Keck Science Center, Claremont Colleges, Claremont, CA 91711-5916 \\
and California Institute of Technology, Pasadena, CA 91125
(Sabbatical)}
\author{R. E. Rothschild}
\affiliation{Center for Astrophysics and Space Sciences,
University of California San Diego, La Jolla, CA 92093}

\date{\today}

\begin{abstract}
Assuming Galactic positrons do not go far before annihilating, a
difference between the observed 511 keV annihilation flux
distribution and that of positron production, expected from
$\beta^+$-decay in Galactic iron nucleosynthesis, was evoked as
evidence of a new source and a signal of dark matter. We show,
however, that the dark matter sources cannot account for the
observed positronium fraction without extensive propagation. Yet
with such propagation, standard nucleosynthetic sources can fully
account for the spatial differences and positronium fraction,
leaving no new signal for dark matter to explain.
\end{abstract}

\pacs{95.35+d, 98.70.Rz, 98.70.Vc}

\maketitle

{\em I. Introduction.---}The positron annihilation radiation in the
511 keV line and 3-photon positronium continuum from the Galactic
bulge has been measured for nearly forty years \cite{jo72,we08} at
a line flux of $\sim 1\times10^{-3}$ photons cm$^{-2}$ s$^{-1}$ and
a continuum flux $\sim$ 3 times that. This corresponds to a
positron annihilation rate in the Galactic bulge of $\sim
1\times10^{43}$ e$^+$ s$^{-1}$. Comparable annihilation radiation
has been measured\cite{we08} from the Galactic disk and halo, which
with their calculated\cite{gu91,je06} positron to 511 keV photon
ratios give annihilation rates of $\sim 0.7\times10^{43}$ e$^+$
s$^{-1}$ for each, assuming a Solar distance to the Galactic center
of 8 kpc.

The long standing model source\cite{co70} of these Galactic
positrons is the $\beta^+$-decay of the radionuclei, $^{56}$Ni,
$^{44}$Ti and $^{26}$Al, in supernova ejecta and Wolf Rayet winds.
They are well defined both theoretically and observationally, and
can easily account for the total Galactic positron annihilation
rate of $(2.4\pm0.2)\times10^{43}$ e$^+$ s$^{-1}$\cite{we08,je06}.

In particular, $^{56}$Ni, which decays through $^{56}$Co to produce
all of the $^{56}$Fe in the Galaxy\cite{ti95}, is by far the
largest single source of Galactic positrons from $\beta^+$-decay of
$^{56}$Co 19\% of the time. That yields about
$(6\pm2)\times10^{44}$ e$^+$ s$^{-1}$, roughly 25 times that needed
to explain the observed annihilation radiation! Because of the
relatively short, $\sim$ 111 days, decay mean life of $^{56}$Co,
most of its $\sim$ 0.63 MeV positrons slow down and annihilate deep
in the expanding supernova ejecta, where their annihilation
radiation is absorbed\cite{co69}. However, an estimated 5$\pm$2\%
of positrons are born late enough to survive annihilation in the
rarifying ejecta of SNIa, based on both model calculations, and
analyses of late-time light curves\cite{ch93}. SNIa produce roughly
half of the Galactic $^{56}$Ni\cite{ti95} with an estimated
Galactic rate of 1 SN every 250$\pm$100 yr, based on extragalactic
surveys\cite{va91} and a $^{56}$Ni yield of 0.58
M$_\odot$/SN\cite{no84}, consistent with their ``standard candle"
luminosity\cite{le01}. This gives a net Galactic positron
production of $(1.6\pm0.6)\times10^{43}$ e$^+$ s$^{-1}$ from the
$^{56}$Ni decay chain, which accounts for roughly 2/3 of the
observed radiation\cite{we08,je06}. This is also quite consistent
with 511 keV line flux limits\cite{ka06} from observations of a
SNIa remnant.

The remaining positrons come from much longer lived $^{44}$Ti and
$^{26}$Al $\beta^+$-decay and essentially all escape from the
ejecta. $^{44}$Ti decays with an 89 yr mean life through $^{44}$Sc
to produce essentially all of the $^{44}$Ca in the Galaxy, and its
production can be directly scaled\cite{ch93} to that of $^{56}$Fe
from the measured Solar system $^{44}$Ca/$^{56}$Fe abundance ratio,
resulting in $(0.5\pm0.2)\times10^{43}$ e$^+$ s$^{-1}$, or 20\% of
that observed. Lastly, the $\beta^+$-decay positrons from $^{26}$Al
with a $1\times10^6$ yr decay mean life are determined directly
from its Galactic mass of $2.8\pm0.8$ M$_\odot$, given by its
measured\cite{ma84} Galactic 1.809 MeV decay line emission. Other
sources, such as cosmic-ray interactions, novae, pulsars, black
holes and gamma-ray bursts, are expected\cite{ra79} to make only
negligible contributions.

The first observations using the spectrometer on the INTEGRAL
satellite\cite{je03} confirmed the earlier 511 keV line flux from
the Galactic bulge at $0.99\times10^{-3}$ photons cm$^{-2}$
s$^{-1}$. However, they found that the bulge-to-disk ratio (B/D) of
the 511 keV positron annihilation line flux from the bulge relative
to that from the disk, exceeded by a factor of 3 or more, the
corresponding ratio of positron production distribution, expected
from the bulge and disk components of SNIae.

If the annihilating positrons do not diffuse far before they slow
down and annihilate, the spatial distribution of their
annihilation should be essentially the same as their production.
Based on that supposition, the INTEGRAL observers suggested
\cite{we04} that some new bulge source of additional positrons
was responsible for the difference in the B/D ratio, although
they had not found any actual increase in the total positron
annihilation from the bulge, compared to previous
measurements\cite{jo72}.

{\em II. Positrons from Dark Matter?---}Dark matter (DM) was
promptly proposed\cite{bo04} as the new source of bulge positrons
and the supposition of negligible positron propagation was
quantified. They suggested that the positrons were formed at
energies of 1 to 100 MeV from $e^+-e^-$ pair-production in the
annihilation of some new light scalar DM particles and
antiparticles. Such positrons, like those from $\beta^+$ decay of
radionuclei, must slow down to nearly thermal energies $\leq$ 10
eV, primarily by ionization losses in the interstellar medium
(ISM), before they in turn can annihilate to produce the observed
511 keV line emission. This slowing-down or stopping distance,
$d_{sd}$, is around $10^{24}$ to $10^{26}$ cm for such positrons in
interstellar gas of density $\sim 0.1$ H cm$^{-3}$. They further
assumed\cite{bo04} that such positrons diffuse through the ISM with
a diffusion mean free path, $\lambda \sim r_L$, the Larmor radius,
of around $10^{9}$ to $10^{11}$ cm in interstellar magnetic fields
of a few microgauss. Therefore, the positrons would only travel a
mean distance $l_{sd} \sim (d_{sd}\lambda)^{1/2} \sim
3\times10^{16}$ to $3 \times10^{18}$ cm, or barely 0.01 to 1 pc,
from their sources before they stopped and annihilated.

Thus, they\cite{bo04} expected the spatial distribution of the
positron annihilation to be virtually the same as that of their
production. Although the DM density distribution in the bulge is
assumed to have a radial power-law form, $R^{-\gamma}$, the
exponent is highly uncertain. So, they simply determined a $\gamma$
of 0.4 to 0.8 from the best-fit Gaussian HWHM $\sim
4.5^{+4.5}_{-1.5}$ deg. of the INTEGRAL angular distribution of the
511 keV line flux\cite{je03}. They also let the observed 511 keV
flux determine an assumed DM-DM* annihilation cross section that
could make it consistent with the assumed relic DM density. Then
with $\sim$10 degrees of freedom, they invented a light DM
candidate that didn't violate collider limits\cite{bo04} and might
be a new source for the presumed bulge positron ``excess".

Based on claims there was no astrophysical explanation, about 150
papers have since been published\cite{bo09}, envisioning a whole
zoo of new DM candidates to explain the bulge ``excess." These
include new axinos, supermassive strangelets, superconducting
strings, Q-balls, sterile neutrinos, mirror matter, moduli,
millicharged fermions, unstable branons, excited WIMPS, electron
interacting $\chi^o$s, etc.\cite{ho04,be08}. Unlike the
astrophysical sources whose positron production and spatial
distributions are determined by independent measurements, all of
these DM candidates are {\em ad hoc}, using the 511 keV flux
measurements to define the needed DM spatial distributions and
decay or annihilation rates.

But can any of these DM candidates really explain the Galactic
bulge positron annihilation radiation? These suggested candidates
ignore the consequences of the fundamental spectral property of the
annihilation emission, that $94\pm4$\% of the
observed\cite{jo72,ch05,je06} annihilation from the bulge occurs
via positronium (Ps) formation, rather than direct annihilation.
Only 25\% of the Ps annihilation occurs in the singlet state,
producing the 2-photon 511 keV line emission, and 75\% occurs in
the triplet state, producing a distinctive 3-photon continuum
feature, while the direct annihilation produces only 2-photon 511
keV line emission. The Ps annihilation fraction can be directly
determined from the observed ratio of the 511 keV line flux to that
of the 3-photon continuum feature. The observed flux ratio
requires\cite{gu91,je06} that the bulge positrons annihilate almost
entirely in the cooler ($< 10^4$ K) HII and HI phases of the ISM,
where Ps formation is $\sim$98\%, rather than in the hot ($>10^6$
K) plasma, where it is only 20\% to 40\%, depending on the dust.

These HI and HII phases in the bulge are primarily confined to the
dense envelopes of molecular clouds, which are concentrated in the
central molecular zone and the surrounding tilted disk. The thin
($\sim$ 0.1 kpc), tilted disk\cite{fe07} is inclined 29$^o$ from
the Galactic plane, and extends $\pm$0.8 kpc, or $\pm5^o$ above and
below the plane and $\pm$1.3 kpc, or $\pm9^o$ in the plane, as
viewed nearly face-on from the Earth. Thus, annihilation in the
tilted disk appears as part of the bulge, rather than disk,
component in the INTEGRAL analyses. As a result, the $\sim 3^o$ to
9$^o$ HWFM of the 511 keV flux reflects this HI and HII gas
distribution, not the positron source.

Yet this HI and HII gas fills barely 1\% of the total interstellar
bulge volume\cite{fe07} within 3 kpc of the Galactic Center. Nearly
all of the suggested DM positrons, however, are expected to be born
throughout the bulge volume over a scale of $\geq$ kpc, so they
would be born almost entirely in the hot plasma, since it fills the
other 99\% of that volume. Therefore, under the assumption that the
positrons do not diffuse more than $\sim$ pc from their points of
origin, most positrons from DM candidates could not get into the HI
and HII gas and could not account for the bulge Ps annihilation
fraction. To account for the observed Ps fraction, thus, requires
positron propagation on scales of $\sim$ kpc.

Only those few scenarios in which positrons are produced from the
interactions of the DM with baryonic or leptonic matter, e.g.
\cite{be08}, could concentrate the positron production in the
molecular clouds to produce the bulge Ps annihilation fraction.
But then the B/D spatial flux ratio could not be satisfied,
because most of the mass of the ISM in is the disk rather than the
bulge\cite{fe01,fe07}.

{\em III. Positron propagation.---}However, the assumption that the
$\lambda$ of 1 to 100 MeV positrons in the general ISM is equal to
$r_L$, and that they therefore travel no more than a pc in their
lifetime\cite{bo04}, is inconsistent with the current understanding
of charged particle propagation\cite{we74} by Larmor resonance
scattering. Diffusion on the scale of the $r_L$ occurs only in
extremely turbulent plasmas, such as relativistic shocks in young
SN remnants\cite{uc07}. But such intense plasma turbulence is
highly localized and very short-lived, and these conditions are not
applicable to the diffuse ISM, where the mean turbulence is far
weaker.

Such a small assumed $\lambda$ in the ISM is also inconsistent with
the direct observations of cosmic-ray electrons and positrons
extending down to this energy range\cite{fa68}, as well as their
synchrotron radio emission from the Galactic halo\cite{sy59}. The
electron $\lambda$ needed to explain these observations has long
been shown\cite{sy59} to be $>$ 1 pc, so they can travel $>$ 1 kpc.

Similar $\lambda$s in the bulge plasma can be understood in terms
of the standard particle propagation theory\cite{we74}. Here
charged particle propagation along magnetic flux tubes is dominated
by resonant pitch angle scattering by cascading MHD waves at the
particle $r_L$. But $\lambda$ is essentially equal to $r_L$ divided
by the scattering probability, which is proportional to the
relative energy density in the MHD waves at $r_L$ compared to that
in the total magnetic field, $\lambda \sim r_L/(\delta
B_L^2/B_o^2).$ That probability is $(\delta
B_t^2/B_o^2)(r_L/l_o)^{2/3}$, the product of the relative magnetic
energy density in turbulence times the relative energy in the MHD
Kolmogorov cascade spectrum at $r_L$ compared to that at the
initial turbulent scale, $l_o$.

In the hot ($\sim10^7$ K), tenuous ($\sim0.005$ H cm$^{-3}$) plasma
with $B_o \sim10 \mu$G fields that fills the $\sim$ 3 kpc
bulge\cite{fe07}, 1 to 10 MeV positrons have a stopping distance,
$d_{sd} \sim$ 3 to 30 Mpc, but $r_L \sim 10^{-10}$ to $10^{-9}$ pc.
However, the relative turbulent field energy density $(\delta
B_t^2/B_o^2) \sim 10^{-2}$ in that phase, assuming it is driven by
the bulge component of SNe\cite{ma04}. Occurring at a rate of
$\sim$ 1SN/1000 yr$^{-1}$ and dissipating $10^{51}$ erg in ejecta
energy over a cascade lifetime of $\sim 10^6$ yr, SNIa generate a
mean bulge turbulent energy density of $\sim 10^{-13}$ erg
cm$^{-3}$ compared to an ambient value of $\sim 10^{-11}$ erg
cm$^{-3}$. With an $l_o \sim$ 50 pc\cite{ya04}, comparable to that
of the SNIa remnants, the relative MHD wave energy at $r_L$
compared to that at $l_o$ is only $(r_L/l_o)^{2/3} \sim
(2-7)\times10^{-8}$. Therefore, the relative energy density in the
MHD waves at $r_L$ compared to that in the total magnetic field,
$(\delta B_L^2/B_o^2) \sim (2-7)\times10^{-10}$, so with $r_L \sim
10^{-10}$ to $10^{-9}$ pc the positron $\lambda \sim r_L/(\delta
B_L^2/B_o^2) \sim$ 0.2 to 5 pc. Thus, such positrons can travel a
mean distance $l_{sd} \sim (d_{sd}\lambda)^{1/2} \sim$ 1 to 10 kpc
from their sources before they will stop and annihilate in the hot
plasma.

Therefore, the bulk of the positrons born in the bulge can diffuse
through the hot plasma without annihilating until they encounter
the HI and HII envelopes of the molecular clouds. In these very
dense envelopes, similar to that of the cloud cores ($\sim$ 1000 H
cm$^{-3}$), the positron $d_{sd}$ is only $\sim$ 10 pc, which is
comparable to the thickness of the envelopes, so the positrons
quickly slow down and annihilate in the cloud envelopes via Ps
formation with the Ps fraction of 94$\pm$4\% \cite{jo72,ch05,je06}.

Far more important, however, we clearly see that the basic
propagation assumption, on which it was claimed that
astrophysical sources failed and DM sources were needed, was
invalid.

{\em IV. Consequences of propagation.---}Thus even though positron
propagation on a kpc scale could solve the Ps fraction problem for
DM positrons, propagation on that scale also easily explains the
difference between the INTEGRAL B/D ratio of positron annihilation
and that of their production by SNIa in the standard model of
Galactic iron nucleosynthesis. That difference, of course, together
with the assumption of negligible propagation, was why it was
argued that there was no astrophysical solution and the only reason
that DM matter solutions were proposed.

In particular, we have recently shown\cite{hi09} that such
propagation not only solves both the fundamental B/D and Ps
fraction problems but also explains many other detailed features of
the Galactic annihilation radiation. Using the SNIa rate based on
stellar mass distributions\cite{de98} and the measured $^{26}$Al
distributions, roughly half of the $\sim 2.4\times10^{43}$ e$^+$
s$^{-1}$ total Galactic $\beta^+$-decay positrons are born within
the interstellar bulge $<$3 kpc. With the current understanding of
propagation\cite{we74}, we show that about 80\% of these positrons
are expected to annihilate via Ps formation in the HI and HII gas
in the bulge, and the remaining 20\% are expected to escape into
the halo. That gives a total bulge production of $\sim
1.0\times10^{43}$ e$^+$ s$^{-1}$, equal to the observed\cite{we08}
bulge annihilation rate. The other half of the $\beta^+$-decay
positrons, produced in the Galactic disk, are born primarily in the
pervasive warm neutral gas\cite{fe01}. There MHD waves are damped
out by ion-neutral friction\cite{hi84} well before they cascade
down to the positron $r_L$, so these positrons stream through the
gas at close to $c$, and only about 50\% of them, or $\sim
0.6\times10^{43}$ e$^+$ s$^{-1}$, are expected to annihilate in the
disk HI and HII gas via Ps, while the other half escape into the
halo. Thus, the expected bulge to disk ratio of positron
annihilation $B/D \sim 1.0/0.6 \sim 1.6$ fully accounts for the
observed\cite{we08} ratio of $1.4\pm0.3$, leaving no unexplained
positron ``excess" in the bulge.

Finally, all of the $\beta^+$-decay positrons that escape from the
bulge and disk, $\sim 0.8\times10^{43}$ e$^+$ s$^{-1}$, naturally
account\cite{hi09} for the Galactic halo annihilation rate of
$\sim 0.7\times10^{43}$ e$^+$ s$^{-1}$, inferred from the
INTEGRAL\cite{we08} 511 keV flux from the halo, adjusted for the
expected\cite{gu91,je06} halo Ps annihilation fraction. So, even
though most of the Galactic DM is assumed to be in the halo, no
unexplained 511 keV flux is observed there either.

{\em V. Conclusions.---}We have shown that the suggested need for
a new source of Galactic bulge positrons to explain the spatial
differences between the observed positron annihilation
distribution and that of $\beta^+$-decay positron production from
Galactic iron nucleosynthesis in SNIa was based only on the
supposition that the positrons do not travel far before they slow
down and annihilate. We showed, however, that the proposed DM
sources can not account for the observed Ps fraction without
extensive propagation. Although we show that such propagation is
expected from current models, we show that then both the spatial
differences\cite{we08,we08b} and the Ps fraction\cite{ch05,je06}
of the Galactic positron annihilation can be fully
explained\cite{hi09} in all of its details by the standard
radionuclei $\beta^+$-decay sources, anchored on Galactic iron
nucleosynthesis. Thus, there is no ``excess" signal left to
explain.
% \vspace*{-12pt}

\end{document}